\documentstyle[epsf]{mn}
 \input epsf.sty
\newif\ifAMStwofonts
\ifoldfss
  \ifCUPmtlplainloaded \else
    \NewTextAlphabet{textbfit} {cmbxti10} {}
    \NewTextAlphabet{textbfss} {cmssbx10} {}
    \NewMathAlphabet{mathbfit} {cmbxti10} {} % for math mode
    \NewMathAlphabet{mathbfss} {cmssbx10} {} %  "   "    "
  \fi
  \ifAMStwofonts
    \ifCUPmtlplainloaded \else
      \NewSymbolFont{upmath} {eurm10}
      \NewSymbolFont{AMSa} {msam10}
      \NewMathSymbol{\upi}     {0}{upmath}{19}
      \NewMathSymbol{\umu}     {0}{upmath}{16}
      \NewMathSymbol{\upartial}{0}{upmath}{40}
      \NewMathSymbol{\leqslant}{3}{AMSa}{36}
      \NewMathSymbol{\geqslant}{3}{AMSa}{3E}

    \fi
  \fi
\fi % End of OFSS

\ifnfssone
  \newmathalphabet{\mathit}
  \addtoversion{normal}{\mathit}{cmr}{m}{it}
  \addtoversion{bold}{\mathit}{cmr}{bx}{it}
  \newmathalphabet{\mathbfit} % math mode version of \textbfit{..}
  \addtoversion{normal}{\mathbfit}{cmr}{bx}{it}
  \addtoversion{bold}{\mathbfit}{cmr}{bx}{it}
  \newmathalphabet{\mathbfss} % math mode version of \textbfss{..}
  \addtoversion{normal}{\mathbfss}{cmss}{bx}{n}
  \addtoversion{bold}{\mathbfss}{cmss}{bx}{n}
  \ifAMStwofonts
    \ifCUPmtlplainloaded \else
      \UseAMStwoboldmath
      \makeatletter
      \new@mathgroup\upmath@group
      \define@mathgroup\mv@normal\upmath@group{eur}{m}{n}
      \define@mathgroup\mv@bold\upmath@group{eur}{b}{n}
      \edef\UPM{\hexnumber\upmath@group}
      \new@mathgroup\amsa@group
      \define@mathgroup\mv@normal\amsa@group{msa}{m}{n}
      \define@mathgroup\mv@bold\amsa@group{msa}{m}{n}
      \edef\AMSa{\hexnumber\amsa@group}
      \makeatother
      \mathchardef\upi="0\UPM19
      \mathchardef\umu="0\UPM16
      \mathchardef\upartial="0\UPM40
      \mathchardef\leqslant="3\AMSa36
      \mathchardef\geqslant="3\AMSa3E
    \fi
  \fi
\fi % End of NFSS release 1

\ifnfsstwo
  \DeclareMathAlphabet{\mathbfit}{OT1}{cmr}{bx}{it}
  \SetMathAlphabet\mathbfit{bold}{OT1}{cmr}{bx}{it}
  \DeclareMathAlphabet{\mathbfss}{OT1}{cmss}{bx}{n}
  \SetMathAlphabet\mathbfss{bold}{OT1}{cmss}{bx}{n}
  \ifAMStwofonts
    \ifCUPmtlplainloaded \else
      \DeclareSymbolFont{UPM}{U}{eur}{m}{n}
      \SetSymbolFont{UPM}{bold}{U}{eur}{b}{n}
      \DeclareSymbolFont{AMSa}{U}{msa}{m}{n}
      \DeclareMathSymbol{\upi}{0}{UPM}{"19}
      \DeclareMathSymbol{\umu}{0}{UPM}{"16}
      \DeclareMathSymbol{\upartial}{0}{UPM}{"40}
      \DeclareMathSymbol{\leqslant}{3}{AMSa}{"36}
      \DeclareMathSymbol{\geqslant}{3}{AMSa}{"3E}
    \fi
  \fi
\fi % End of NFSS release 2

\ifCUPmtlplainloaded \else
  \ifAMStwofonts \else % If no AMS fonts
    \def\upi{\pi}
    \def\umu{\mu}
    \def\upartial{\partial}
  \fi
\fi

\title[V1425 Aql - Intermediate Polar Candidate]
  {Nova V1425 Aquilae 1995 - The Early Appearance of Accretion Processes
    in An Intermediate Polar Candidate}

\author[A. Retter, E.M. Leibowitz and O. Kovo-Kariti]
   {A.~Retter$^*$,
   E.M.~Leibowitz
   and O.~Kovo-Kariti\\
  School of Physics and Astronomy and the Wise Observatory,
Raymond and Beverly Sackler Faculty of Exact Sciences,\\
Tel-Aviv University, Tel Aviv, 69978, Israel\\
   $^*$ email: alon@wise.tau.ac.il\\}

\date{submitted 1997 April 10}
\pagerange{\pageref{firstpage}--\pageref{lastpage}}
\pubyear{1997}

\def\LaTeX{L\kern-.36em\raise.3ex\hbox{a}\kern-.15em
    T\kern-.1667em\lower.7ex\hbox{E}\kern-.125emX}

\begin{document}

\label{firstpage}

\maketitle

\begin{abstract}

Continuous CCD photometry of Nova Aquilae 1995 was performed through
the standard $B,V,R$ and $I$ filters during three nights in 1995 and
with the $I$ filter during 18 nights in 1996. The power spectrum of the
1996 data reveals three periodicities in the light curve: 0.2558 d,
0.06005 d and 0.079 d, with peak-to-peak amplitudes of about 0.012,
0.014 and 0.007 mag. respectively.

The two shorter periods are absent from the power spectrum of the 1995
light curve, while the long one is probably already present in the
light curve of that year.

We propose that V1425 Aql should be classified as an Intermediate -
Polar CV.  Accordingly the three periods are interpreted as the orbital
period of the underlying binary system, the spin period of the magnetic
white dwarf and the beat period between them. Our results suggest that
no later than 15 months after the outburst of the nova, accretion
processes are taking place in this stellar system. Matter is being
transferred from the cool component, most likely through an accretion
disc and via accretion columns on to the magnetic poles of the hot
component.

 \end{abstract}


 \section{Introduction}

 \subsection{V1425 Aquilae}

Nova Aql 1995 was discovered on 1995 February 7 by Takamizawa
(1995). Mason et al. (1996) used its similarity to Nova V1668 Cygni
1978 to extrapolate a maximum brightness of $M_{V}\approx 6.2$ and
$t_{2V}\approx 11$ d. This classifies Nova Aquilae as a fast one. They
also deduces from IR considerations that the dust shell of the nova was
optically thin shortly after its outburst. The probable precursor star
of about 20 mag. was detected by Skiff (1995) and this indicates an
outburst amplitude of about 14 mag.

Retter, Leibowitz \& Kovo-Kariti (1996) and Leibowitz, Retter \&
Kovo-Kariti (1997) reported the discovery of two photometric periods in
the light curve of Nova Aquilae 1995, and pointed out the resemblance
of this object to Intermediate Polars systems. In this work we present
further arguments for our interpretation of the periods of V1425 Aql as
resulting from the Intermediate Polar nature of the nova.

\subsection{Intermediate Polars and nova systems}

Intermediate polars (for reviews see Patterson 1994, Hellier 1995,
1996, 1997) are binary systems, which are a sub-class of AM Her stars
(magnetic Cataclysmic Variables). Unlike in other members of the AM Her
class, in Intermediate Polars the rotation of the primary white dwarf
is not synchronized with the orbital motion of the binary system. The
spin periods found in Intermediate Polars are usually much shorter than
their orbital periods (Patterson 1994, Hellier 1996).

One of the main observational characteristics of Intermediate Polars is
the presence of multiple periodicities in the power spectra of their
light curves, emanating from the non-synchronous rotation of the
primary with the orbital revolution.  In fact this characteristic has
become, together with modulations of the X-ray radiation of the system,
a major criterion for membership in the Intermediate Polar group.

The classification of a classical nova as an Intermediate Polar is not
new. DQ Her is a prototype of a nova which is an Intermediate Polar
(Patterson 1994), and in the last few years a few other novae have been
so classified as well. Examples are V533 Her and GK Per, with white
dwarf spin periods of 64 and 351 sec, respectively (Patterson 1994).
The detected variations of V603 Aql in the optical, X-ray and
ultraviolet regimes make it also a likely Intermediate Polar system
(Udalski \& Schwarzenberg-Czeny 1989, Schwarzenberg-Czeny, Udalski \&
Monier 1992). The presence of multiple periods in the light curve of HZ
Pup supports the membership of this nova in the Intermediate Polar
group, too (Abbott \& Shafter 1997). In V1974 Cygni, evidence was found
to the presence of an accretion disc in the system, (Retter, Ofek \&
Leibowitz 1995, Retter, Leibowitz \& Ofek 1996, 1997a, 1997b, 1997c,
Skillman et al.  1997), together with some indications for an intense
magnetic field on the surface of the white dwarf (Chochol et al.
1997).  The combination of these two properties makes V1974 Cyg another
potential candidate for the Intermediate Polar group.

It would therefore appear that the Intermediate Polar phenomenon is not
uncommon among classical nova CV systems.

\section{Observations}

We observed Nova Aquilae during three nights in 1995 May, and 18 nights
during the interval 1996 April -- August. Table 1 presents a summary of
the observations schedule. We used the Tecktronix 1K CCD camera,
described in Kaspi et al. (1995) mounted on the 1-m telescope at the
Wise Observatory. During the observations in 1995 we switched
successively between the standard $B,V,R$ and $I$ filters, and in 1996
the photometry was carried out only in the $I$ band. We note that our
$I$ filter is slightly redder than the standard bandpass. The exposures
times of the observations in 1995 were 30 sec. ($I$), 40 ($R$), 50 ($V$)
and 60 ($B$) with a repetition time of about 250 sec. In 1996 the
integration time was 180 sec. The number of frames obtained in our
programme in 1995 and 1996 are 522 and 1170, respectively.

\begin{table}
  \caption{The observations time table}
  \begin{tabular}{@{}ccccc@{}}

UT&Time of Start&Run Time&Points&Filters\\
   Date  &(JD+2449000) & (hours)&number&\\
\\

100595& 848.410&       4.5 &146&B,V,R,I\\
110595& 849.403&       4.5 &202&B,V,R,I\\
120595& 850.408&       4.5 &174&B,V,R,I\\
280496& 1202.497&      2.3 &41&I\\
290496& 1203.496&      1.3 &17&I\\
100596& 1214.493&      1.9 &32&I\\
110596& 1215.637&      0.2 &4&I\\
300596& 1234.419&      3.8 &59&I\\
310596& 1235.373&      2.6 &47&I\\
010696& 1236.393&      4.5 &82&I\\
310796& 1296.256&      5.9 &103&I\\
010896& 1297.234&      6.7 &120&I\\
020896& 1298.252&      6.4 &96&I\\
030896& 1299.238&      7.2 &115&I\\
210896& 1317.251&      4.5 &80&I\\
220896& 1318.238&      4.0 &70&I\\
230896& 1319.283&      4.0 &35&I\\
270896& 1323.234&      4.3 &62&I\\
290896& 1325.244&      4.3 &79&I\\
300896& 1326.224&      4.2 &70&I\\
310896& 1327.245&      4.3 &58&I\\

\end{tabular}
\end{table}

Photometric measurements were performed using the DAOPHOT program
(Stetson 1987). An IRAF\footnote{IRAF (Image Reduction and Analysis
Facility) is distributed by the National Optical Astronomy
Observatories, which are operated by AURA, Inc., under cooperative
agreement with the National Science Foundation.} script was written by
the first author for automatic reduction with aperture photometry. We
chose star radii of five pixels, corresponding to angular diameter of
about 3.5 arcseconds. Instrumental magnitude of the nova, as well as of
a few dozens reference stars, depending on the image quality, were
obtained from each frame.  We finally used the Wise Observatory
reduction program DAOSTAT (Netzer et al. 1996) for an internal
consistent series of the nova magnitudes.

Fig. 1 displays a comprehensive visual light curve of the nova from
discovery to 1996 September. The data were taken from
AFOEV\footnote{AFOEV (Association Francaise des Observateurs d'Etoiles
Variables) operates at Strasbourg Astronomical Data Center (CDS),
France.}. The marks in the figure indicate the times of our
observations.

\begin{figure}

\centerline{\epsfxsize=3.0in\epsfbox{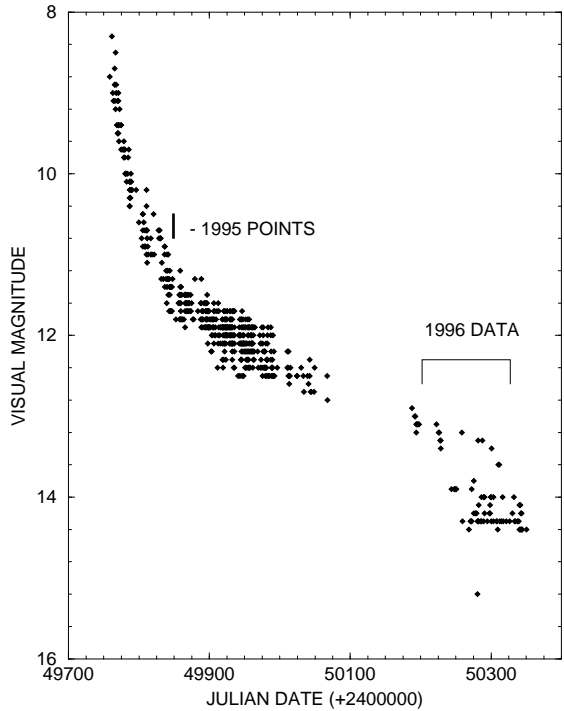}}

\caption{Two years light curve of Nova~Aql~1995. Data points are visual
estimates of amateur astronomers, compiled by AFOEV$^\dagger$. The
times of our observations are marked.}

\end{figure}

Fig. 2 presents the 1996 $I$ light curve of the nova, as measured in
our observing programme. One can see that during the time interval
spanned by the observations in 1996, the nova declined by about 0.6
mag. A much faster decrease in the brightness of the system (almost
four mag.) describes the behaviour of the nova in the $I$ band from
1995 May to 1996 April.

\begin{figure}

\centerline{\epsfxsize=3.0in\epsfbox{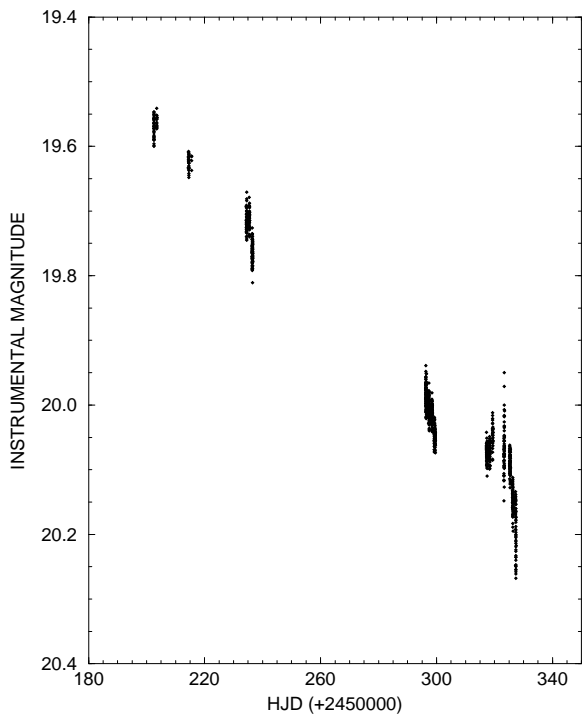}}

\caption{All the 1996 $I$ magnitudes of V1425 Aql that are presented
in this work. The brightness of the nova decreased by about 0.6 mag.
during the time interval of these observations.}

\end{figure}

 \section{Data Analysis}

 \subsection{The apparent periodicities}

During a few of our best nights in 1996, the light curve of the nova
exhibits a small sinusoidal-like modulation with a peak-to-peak
amplitude of about 1 -- 2 per cent on a time scale of 1.5 hr. Sometimes
a smaller dip at phase 0.5, relative to the main variation, was also
seen. However, during most of the nights these variations were hidden
in the noise. In a few nights a systematic brightness variation of a
time scale of a few hours, with similar amplitude, could be recognized
in the light curve as well.

The upper panel of Fig. 3 is a plot of the power spectrum (Scargle
1982) of most of the $I$ band points in 1996. We omitted from our
analysis the night of May 11, because it had only a very few points,
and the data points of August 27 and 31, when the light of the nova and
the comparison stars varied by about 0.1 mag, due to reflected
moonlight or instrumental problems. The inclusion of these three nights
in the analysis doesn't affect the results in any significant way.
Altogether we used 1046 points in our time series analysis.  Before
applying the power spectrum routines we normalized the data by
subtracting the mean magnitude from each night.

The power spectrum shows three distinct groups of peaks. The left one
is around the frequency 3.909 d$^{-1}$, corresponding to the period
P$_{1}=$ $0.2558\pm0.0001$ d. The uncertainty interval is a better than
99\% confidence limit for the value of the period. It is derived with
the Bootstrap technique (Efron \& Tibshirani 1993) from a sample of
1000 pseudo-observed light curves.  The second group near the center of
the frame is around the frequency 16.6518 d$^{-1}$, corresponding to
the period P$_{2}=0.06005\pm0.00001$ d, with the uncertainty derived as
for the period P$_{1}$. The third group of peaks at the right-hand side
of the diagram is centered around the first harmonic of P$_{2}$. All
three groups are characterized by a central frequency and a pattern of
peaks on both sides, corresponding to the 1,1/2,1/3 etc. day aliases of
the central one.

\begin{figure}

\centerline{\epsfxsize=3.0in\epsfbox{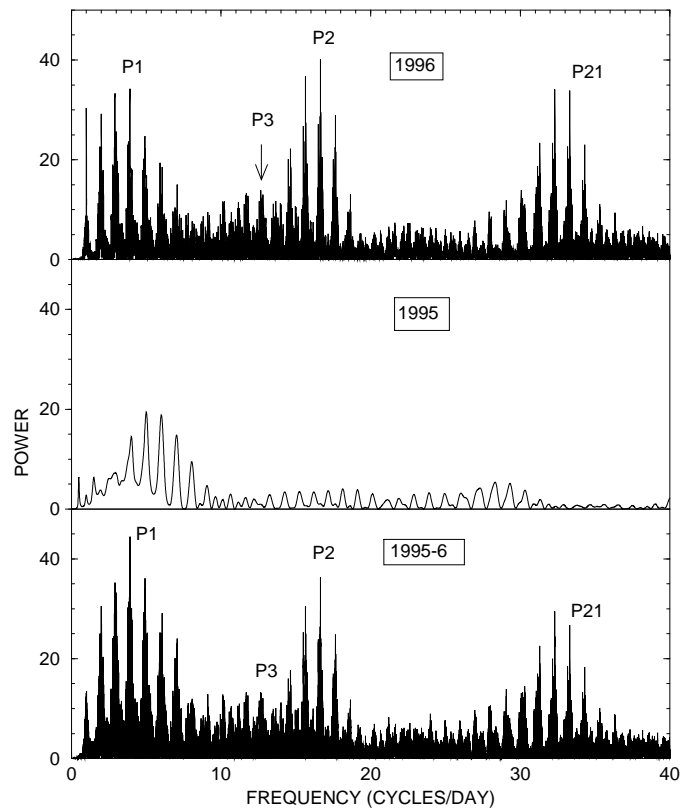}}

\caption{{\bf Upper panel} - power spectrum of the $I$ light curve of
15 nights in 1996. The peaks marked as P1 and P2 represent the two
major periodicities in the light curve, while peak P21 is the first
overtone of P2. Each one is at a center of a pattern of peaks
corresponding to the 1, 1/2, 1/3 etc. day aliases of the central one.
The peak P3 is at the beat frequency between P1 and P2.
{\bf Middle panel} - power spectrum of the three nights in 1995. The
data are a mixture of measurements in four bands. Within the large
observational uncertainty, the group of high peaks at the left-hand
side of the diagram (being aliases of each other) may be regarded as
representing the same periodicity associated with the P1 peak in the
1996 power spectrum.
{\bf Lower panel} - power spectrum of the combined 1995-6 data. The P1
peak is significantly higher than in the 1996 data, while the other
peaks are somewhat lower.}

\end{figure}

We checked the reliability of the three major periodicities in the
light curve of the nova by dividing the data into two distinct parts
(the first eight nights, and the remaining seven nights). A similar
triple structure appeared in the power spectrum of each part
separately. In a different test for the independence of the three
periodicities, we subtracted from the data the fundamental harmonic of
the lower frequency. The other two peaks remained almost unaltered in
the power spectrum of the residuals. We repeated this treatment for the
other two frequencies and found out that all three are indeed
independent of each other. As a further check we created an artificial
light curve on the times of the real observations, by superposing a
sine wave of one of the three periodicities over a random distribution
of points representing white noise. The power spectrum of each of these
three synthetic light curves showed only the corresponding planted
periodicity, surrounded by an alias pattern similar to the one of the
real data, with no trace of the other periodicities.

\subsection{A third Periodicity}

In the power spectrum shown in the upper panel of Fig. 3 there is an
additional group of peaks that stand out considerably above the noise
level, although not to the height that makes them statistically
significance.  The central highest ones are at the frequency 12.70
d$^{-1}$, corresponding to the period P$_{3}=0.079$ d, and its one day
alias at the frequency 11.72 d$^{-1}$. If we remove from the data the
two periods P$_{1}$ and P$_{2}$, represented by the fundamental and the
first harmonic of the corresponding periodicities, the group of these
peaks remains almost unaltered in the power spectrum of the residuals.

The frequency of the beat between the P$_{1}$ and the P$_{2}$
periodicities is 12.74 d$^{-1}$, corresponding to the period 0.0785 d.
The near similarity between this frequency and the P$_{3}$ frequency in
the power spectrum of the observed data strongly suggests the presence
of the beat period in the 1996 light curve of the nova, in addition to
the presence of the periods themselves. The fact that the beat
frequency remains present in the power spectrum even after the removal
of the other two periodicities indicates that this frequency is indeed
modulating the nova light and that it is not a numerical artifact of
the data reduction of the very unevenly sampled light curve of the
system.

We can estimate the statistical significance of the P$_{3}$ peak in the
power spectrum in the following way. By the Bootstrap technique on a
sample of 1000 pseudo-light curves, we found that a one Sigma (67\%)
and a 99\% uncertainty intervals in the position of the P$_{3}$ in the
power spectrum are 0.0004 and 0.001 d$^{-1}$, respectively. Thus it
appears that the observed P$_{3}$ peak falls within one Sigma of the
expected position of the beat frequency.  Applying now again the
Bootstrap routine on another sample of 1000, we found that the
probability of a chance occurrence of a peak as high as the observed
one within a 99\% uncertainty interval around the beat frequency is
less than 0.1\%.  Thus the combination of the height of the P$_{3}$
peak and the proximity of its position to that of the beat frequency,
whose value is known a-priory, once the P$_{1}$ and the P$_{2}$ periods
are known, makes the P$_{3}$ peak in the power spectrum a statistically
highly significant feature. We therefore conclude that there exists a
true third periodicity, the beat of the first two major ones, in the
light curve of Nova Aql 95 during the 1996 observations.

\subsection{Structure of the three periodicities}

In the three panels of Fig. 4 we show the $I$ band data of the 1996
observations folded on to the periods P$_{1}, P_{2}$ and the beat
period between them, P$_{3}$. The points are the average magnitude value
in each of 40 equal bins that cover the 0~-~1 phase interval. The bars
are the $1 \sigma$ uncertainties in the value of the average values.
Solid line represents the first two terms in a Fourier expansion around
the corresponding fundamental periods, fitted to the data points by
Least Squares. The full amplitude of the average variation in the upper
curve is 0.012 mag., that of P$_{2}$ is 0.014 mag. and that of the
third periodicity at the lower panel is 0.007 mag. These numbers were
derived by binning the folded data into 10 equal intervals and by
measuring the difference between the extrema. The error in the
amplitude in all three cases is about 0.004, and it was calculated by
samples of 1000 Bootstrap simulations.  The double structure in the
P$_{2}$ curve is responsible for the strong appearance of the first
overtone of this periodicity in the power spectrum (Fig. 3 upper
panel). It will be further discussed in section 4.

\begin{figure}

\centerline{\epsfxsize=3.0in\epsfbox{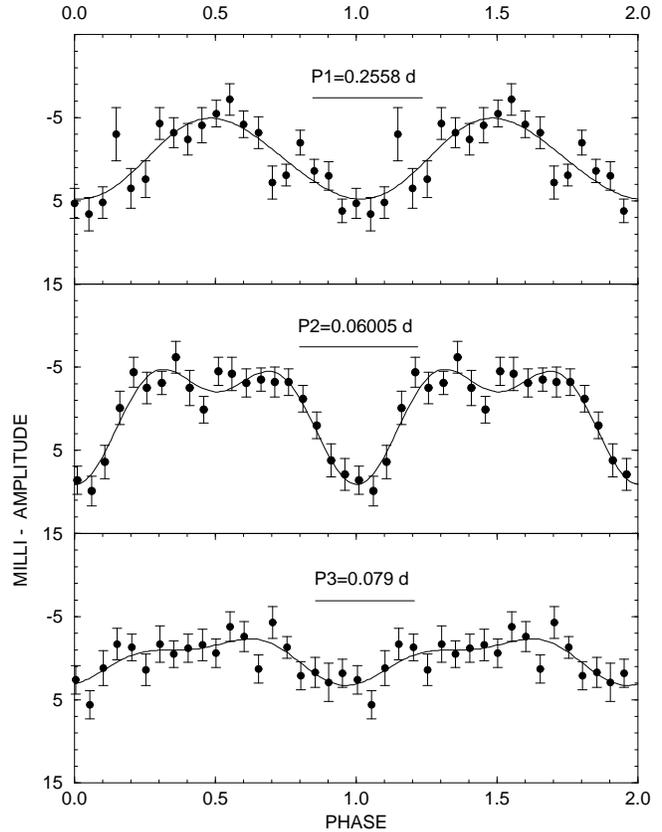}}

\caption{The $I$ filter light curve of 15 nights in 1996, folded on to
the three periods and binned into 20 equal bins. {\bf Upper panel} ---
the 6.14-h period. {\bf Middle panel} --- the 1.44-h periodicity. {\bf
Lower panel} --- the 1.9-h beat period.}

\end{figure}

The best fitted ephemeris for the three periodicities are:\\
\\
$T_{1}(min)$ = HJD 2450202.566 + 0.2558 E.\\
\hspace*{1.23in}$\pm $0.002 \hspace*{0.01in} $\pm$ 0.0001\\
%\hspace*{1.1in}$\pm $0.002 \hspace*{0.01in} $\pm$ 0.0001\\
\\
$T_{2}(min)$ = HJD 2450202.664 + 0.06005 E.\\
\hspace*{1.23in}$\pm $0.007 \hspace*{0.01in} $\pm$ 0.00001\\
\\
$T_{3}(min)$ = HJD 2450202.44 + 0.079 E.\\
\hspace*{1.23in}$\pm $0.01 \hspace*{0.01in} $\pm$ 0.001\\

\subsection{The 1995 light curve}

In 1995 we observed the nova in only three nights, each run lasting
about 4.5 hr. In these observations we used the four standard filters -
$B,V,R$ and $I$ in a sequence. Inspecting the four light curves by eye
we find no clear short scale variations of the order of 1.5 hr. The
light curves do show, however, a long term variation, on time scale of
the duration of the runs, i.e. of a few hours. The four power spectra
of the bands (not shown here) have a few peaks, corresponding to
periods of a few hours, all of them are, however, statistically nearly
insignificant.

In order to improve the significance of the peaks in the power
spectrum, we combined the four bands together and formed a united
colour power spectrum. This was done by subtracting the mean from the
points of each filter at the three nights, and then sorting the data by
time. The power spectrum of the resulting light curve is shown at the
middle panel of Fig. 3. The pattern of 1-d alias peaks at the left-hand
side of the diagram, which is many $\sigma$ above the noise level,
includes the frequency $4.01\pm 0.17$ d$^{-1}$. Due to the scarcity of
the 1995 data, the small amplitude of the variation (0.010 $\pm$ 0.006
mag.) and the rather short duration of the observations in each night,
the uncertainty in the position of this pattern of peaks on the
frequency axis is large.  Within this uncertainty it is consistent with
the lower frequency peak (P$_{1}$) in the power spectrum of the 1996
data. In the 1995 power spectrum, however, there is no trace of the
shorter periodicities identified in the 1996 data.

With the Bootstrap technique we found on a sample of 1000 pseudo-light
curves, that the probability to obtain anywhere in the power spectrum a
pattern of peaks as high as the observed ones is smaller than 0.1\%. As
a more severe test we checked also the probability to obtain peaks as
high as the observed ones in data of correlated magnitudes. For that
purpose we planted on the times of our real observations in each night
a sinusoidal modulation with periods randomly chosen between 4--8 hr.
This test shows that the observed pattern of peaks is significant at a
99\% confidence level.

We also combined the data of 1995 with the 1996 points. In the power
spectrum of the combined light curve (Fig. 3 lower panel), the peaks
around P$_{1}$ gain a considerable amount of power. In particular the
power at this frequency increases from $\sim35$ in the 1996 data to
$\sim45$ in the combined 1995/96 data. This increase should be compared
with a small decrease in the power in the other periodicities of 1996.
For comparison we also combined with the 1996 data a pseudo 1995 light
curve, namely, random magnitudes, distributed over the times of
observations in 1995.  No increase in power at the P$_{1}$ frequency is
obtained in the power spectrum of this light curve, nor in the power
spectrum of a light curve in which we added to the 1996 data a noisy
Sine wave of other periods that is planted on the 1995 times of
observation.

We may summarize that while the 1995 observations are too scarce for
establishing by themselves the presence of a few hours periodicity in
the light curve, much less for determining its value, the data do
suggest that the 6.14-h period identified in the 1996 light curve had
already existed in the 1995 light curve. The 1.44-h periodicity and its
first harmonic, as well as the beat period, do not seem to be present
in the 1995 data.

\section{Discussion}

\subsection{Identification of the two periods}

Three periodicities have been identified in the light curve of V1425
Aql 15 months after its outburst; one of them, P$_{1}$, was present in
the light curve already three months after outburst. The three periods
reflect three genuine modulations of the light emanating from this
stellar system.  They are not independent of each other as one of them
is the beat between the other two. Thus it appears that in 1996 two
independent clocks are operating in the nova system, each one modulates
the light radiation directly and also indirectly through some
combination with the other clock.

We suggest that the longest periodicity, P$_{1}\sim$ 6.14 h, is the
orbital period of the nova underlying binary system. This is based
mainly on the fact that this period was present in the light curve
already in 1995, with no apparent change in its value. The observations
by Mason et al. (1996), and their conclusion, that the dust shell of
the nova was not optically thick at the time of their and our
observations, make it indeed likely that the binary system could have
been seen with optical light at that time. Radial velocity measurements
should confirm or refute our suggestion.

The amplitude of the 6.14-h variation, if indeed present in the 1995
light curve, is comparable, in magnitude units, to the amplitude of
this variation in 1996.  In 1996, however, the optical brightness of
the system was some four mag. fainter than in 1995. This seems to us to
suggest that the major source of the binary modulation in the light
curve is the reflection effect. The amplitude of modulations by this
effect depends on the inclination angle of the system and on the
fraction of the radiation from the hot component, that is being
intercepted by the companion, and reflected in the direction of the
observer.  The first parameter is clearly an invariant of the system.
The second one may also be constant to first order, provided that the
dimensions of the hot component did not change appreciably, relative to
the radius of the companion, between the two years. This could be the
case either because the size of the hot source, indeed did not change
by a large factor from 1995 to 1996, or because in 1995, the size
of the hot source, e.g. the white dwarf or the pseudo-photosphere of
the white dwarf, was already small relative to the radius of the
secondary star.

We further propose that Nova Aql 95 is an Intermediate Polar, in which
the shorter period, P$_{2}$, is the spin period of the magnetic white
dwarf in the system. Spin period that is shorter than the orbital is
the rule in almost all known Intermediate Polars. There is only one
exception, RX J19402-1025, but this is a nearly synchronous system, in
which the spin period has only a marginal excess: $(P_{spin} -
P_{orbital}) / P_{orbital}\sim 0.3$\% (Patterson et al. 1995).

The double structure shape of the 86.5-m period (Fig. 4 middle panel)
is typical to the variations emanating from one pole of a rotating
magnetic white dwarf (Warner 1995). This evidence supports the
interpretation of this period as the spin period of the white dwarf.

Finally, the third period is interpreted as the beat period between the
two other periods. Its independent presence in the light curve of
Intermediate Polar system is well established, observationally as well
as in theory (Patterson 1994, Hellier 1995, 1996, 1997).

\subsection{The accretion form -- an Accretion Disc (?)}

The fate of the accretion disc in a classical nova binary system during
the outburst event and immediately following it is unclear. It is
sometimes assumed that if an accretion disc is present in the
pre-outburst system, it is being destroyed by this cataclysmic event.
However, no theoretical effort was done in the direction of answering
the question when is the accretion disc rebuilt in the shattered,
slowly decaying system.  Up to recent times there were little
observational data concerning the existence of the accretion disc in
young novae.  In the last few years, however, observational data are
being accumulated, indicating an early presence of an accretion disc in
a few classical novae, already a few weeks or months after the
outburst. Leibowitz et al. (1992) discovered an eclipse three weeks
after maximum light in Nova V838 Herculis 1991.  They interpreted it as
the occultation of the accretion disc by the secondary star.  Some 30
months after the outburst of the classical nova V1974 Cyg, Retter,
Leibowitz \& Ofek (1997a, 1997b, 1997c) and Skillman et al. (1997)
detected permanent superhumps in the light curve of this system.
Superhumps characterize the SU UMa class of CVs that are known to have
an accretion disc in their underlying stellar system.  Thus the
observations in V1974 Cyg indicate the early presence of an accretion
disc also in that system.

It is now believed that in nearly all known Intermediate Polars, a
major part of the accretion proceeds through an accretion disc most of
the time (Patterson 1994, Hellier 1996). In a few systems the
accretion is partly maintained by a different mode - disc overflow, in
which the accretion stream from the companion bypasses the accretion
disc and interacts directly with the magnetic field of the white dwarf.
(Hellier 1993, 1995, 1996, 1997, Hellier \& Livio 1994). However,
only one object (RX J1712-24) out of the 13 Intermediate Polars listed
by Hellier 1996  (see his Fig. 2) is believed to be a disc-less
system. Based on this statistics, and ignoring the possibility that it
is biased by a selection effect due to the excessive brightness of the
disc, we may regard it as very likely that no later than 1996 May, Nova
Aql 95 already contained an accretion disc within its binary system.

Unless the asymmetries in the system are large, the relative amplitude
of the spin period to the beat period is a first order measure of the
rate of accretion through an accretion disc relative to the rate of
accretion via the disc overflow mode (Hellier 1997). In N. Aql 95, if
the 86.5-m period is indeed the spin period of the white dwarf, the
dominant accretion is via the accretion disc, while a smaller part of
it is maintained through the disc overflow mode. This is implied by the
(peak-to-peak) amplitude 0.007 mag. of the beat period 0.079 d, that is
a half of the amplitude 0.014 of the 0.06005-d spin period.

The 86.5-m spin periodicity in V1425 Aql is one of the longest among
Intermediate Polars (Patterson 1994, Hellier 1996), especially if the
three nearly synchronous systems (Nova V1500 Cygni 1975, BY Cam and RX
J19402-1025 - Patterson et al. 1995) are not counted in this class
(Warner 1995 groups them with the AM Her systems). Hellier (1996) and
Allan et al. (1996) speculated that slow rotators accrete through
accretion curtains. A 86.5-m spin cycle makes Nova Aquilae 95 a slow
rotator and therefore a system with that type of accretion mode.  In
the accretion curtains model, the pulse structure is expected to
consist of a single hump, because the two poles in this mode, act in
phase.  In the middle panel of Fig. 4 we see, however, that the pulse
in N. Aql 95 has a structure of mid-way between one and two humps.
According to Hellier, this would indicate that in this system polecaps
are modulating the optical radiation to a large extend, in spite of the
slow rotation of the white dwarf.  If these ideas are correct, it is
another evidence for the presence of an accretion disc in the system.

According to the spin-amplitude relation of Patterson (1994) the
amplitude of the shorter, spin variation should increase as the nova
continues to fade. Further confirmation for the Intermediate Polar
nature of this system may come also from future X-ray and polarization
measurements.

\subsection{Magnetic Novae and Speed Class}

It was proposed by Diaz \& Steiner (1991) and by Orio, Trussoni \&
Ogelman (1992) that classical novae with strong magnetic fields tend to
be of higher speed class. Our findings that Nova Aql 95 is an
Intermediate Polar seems to be in line with this suggestion as this
nova is a fast one with $t_{2V}\approx 11$ d (Mason et al. 1996).

We checked this claim against observational data concerning some 60
classical novae (Warner 1995), of which 12 are Intermediate Polars or
suspected Intermediate Polars. We found that the histograms of both
t$_{2}$ and t$_{3}$ parameters of the population of Intermediate Polars
novae are not significantly different from the corresponding histograms
for the non-magnetic systems. Our test, however, cannot be considered
as evidence against the above contention, since the statistic is rather
poor due to the still small number of recognized magnetic classical
novae.

\section{Summary}

(1) We found three periods in the power spectrum of Nova Aql 1995:
0.2558 d, 0.06005 d (with its first harmonic) and 0.079 d - the beat
period between the first two periods.\\
(2) The longer period is seen in the power spectra of the observations 
in the two years 1995 and 1996,
while the other two periodicities are absent from the 1995 data.\\
(3) The inter-relations among the frequencies of the three periods
are characteristics of Intermediate Polar systems. Based on this we
suggest that V1425 Aquilae belongs to this group.\\
(4) We interpret the longer 6.14-h period as the orbital period of the
binary system, the 86.5-m period as the spin period of a magnetic white
dwarf and the 1.9-h period as the beat period between them.\\
(5) These results imply that accretion (probably via an accretion disc)
was active in the system already some 15 months after the outburst of the
nova.

\section{Acknoledgements}

We thank John Dan and the Wise Observatory staff for their assistance
with the observations. We would also like to thank Shai Kaspi for
taking for us a comparable spectrum of the nova.

This research has made use of the AFOEV database, operated at CDS,
France.

Astronomy at the Wise Observatory is supported by grants from the
Israeli Academy of science.

\end{document}